# Liquid Cooling System for a High Power, Medium Frequency, and Medium Voltage Isolated Power Converter

Hooman Taghavi
Department of Mechanical Engineering
University of South Carolina
Columbia, SC, USA
htaghavi@email.sc.edu

Ahmad El Shafei
Department of Electrical Engineering
University of Wisconsin Milwaukee
Milwaukee, Wisconsin, USA
aie@uwm.edu

Adel Nasiri
Department of Electrical Engineering
University of South Carolina
Columbia, SC, USA
nasiri@mailbox.sc.edu

*Abstract*— Power electronics systems, widely used in various applications such as industrial automation, electric cars, and renewable energy, have the primary function of converting and controlling electrical power to the desired type of load. Despite their reliability and efficiency, power losses in these systems generate significant heat that must be dissipated to maintain performance and prevent damage. Cooling systems play a crucial role in ensuring safe operating temperatures for system components. Air and liquid cooling are the leading technologies used in the power electronics world. Air cooling is simple and cost-effective but is limited by ambient temperature and component thermal resistance. While more efficient, liquid cooling requires more maintenance and has higher upfront costs. Water-cooling systems have become famous for regulating thermal loads as they can effectively remove heat from localized high-temperature areas, such as the challenging hotspots in power electronics systems. In addition to designing a cooling system for a power electronic system, this study investigated the impact of three major parameters; cold plate material, channel shape/size, and coolant inlet velocity. The research examined and analyzed these factors and their trade-off analysis to obtain cooling system design and optimization insights. This study might improve power electronics system performance, reliability, and durability by improving heat dissipation and thermal management.

## I. Introduction

Smart Grid (SG) is an advanced power grid technology where bi-directional electricity and information flow create a decentralized automated energy distribution network [1]. It addresses the drawbacks of traditional grids by integrating renewable energy sources, ensuring efficient energy management, and leveraging current information technology [2-5]. With the help of modern power electronics technology, SG can respond to electric power generation, transmission, distribution, and consumption by implementing the necessary strategies. For example, SG can quickly adjust power flow and restore the power delivery service in the event of a breakdown in a distribution grid medium voltage transformer [6].

Energy routers in SG allow consumers to sell surplus electricity back to the utility and manage power flow. Ride-through, fault current limits, and power factor correction improve distribution power quality. SST is a significant and crucial part of SG topologies. Unlike traditional bulky transformers, it can operate in both AC and DC grids and be modularly constructed for bi-directional power transmission [7]. For instance, the SST concept helps in power factor correction and harmonic compensation when the Unified Power Quality Conditioner (UPQC) fails [8].

One of the major SST components that helps connect variable renewable energy sources to the grid is Medium Frequency (MF) DC-DC converters with Medium Frequency Transformer (MFT) utilization. MF power conversion systems in the kilo-hertz band are lighter and smaller than old power systems that use Line Frequency Transformers (LFTs) [8, 9]. LFT's mature technology has long been utilized in traditional utility distribution systems; however, high-power, high-voltage MFTs are not yet mature and standardized. Electromagnetic modeling, insulation coordination, electrical parameter control, thermal coordination, and design optimization need more research and testing before utilities can transition and adopt them for medium voltage renewable energy integration [2].

Higher power, voltage, and frequency levels allow renewable energies and power electronics systems to be used at a larger scale and be integrated with the utility grid. Increasing switching operating frequency makes magnetic components in renewable energy systems go smaller in volume and mass, leading to a higher power density [10]. However, thermal stress and smaller dimensions with high power density hinder MFT design and development, making thermal management and cooling design extremely challenging [9]. High-power SST applications often cascade DC-DC converters to the Medium Voltage (MV) grid. Operating at MV levels presents an additional challenge and trade-off with thermal management to maintain a high-power density requirement [11].

MFT design has been studied for traction, renewable energy, and SGs applications [12, 13]. For instance, an amorphous alloy MFT design is described in [14], and a simplified thermal model optimizes heat dissipation capacity [15]. Another work examines core materials and optimizes temperature. These optimization approaches ignore electric field distribution and insulation requirements, which are crucial for insulation design at medium/high voltage utility grid connection levels. SSTs must address high power, high voltage, and high-frequency requirements and challenges, specifically thermal management and cooling, which is the focus of this paper [16].

## II. DC-DC CONVERTER

DC-DC converters are essential power electronics components for voltage conversion in various applications [17]. These converters are widely used in electrical equipment and motor vehicles to effectively control and distribute electric power [18]. There are different types of DC-DC converters available:

- Buck converters reduce input voltage to a lower output voltage.
- Boost converters increase input voltage to a higher output voltage.
- Buck-boost converters can change the input voltage to a different output voltage.
- Flyback converters generate an isolated output voltage from an input voltage [18].

DC-DC converters are used to step up or down the output voltage based on specific requirements, ranging from small electrical devices to transmission lines [17]. They are extensively used in regulated Switch Mode Power Supply (SMPS) and DC motor drive applications. The duty cycle of the switches, controlled by high-frequency switches like MOSFETs, determines the output voltage adjustment capability of the converter [15].

### A. Dual Active Bridge

The Dual Active Bridge (DAB) is a converter used to connect two DC power sources to a single load. It is commonly used in MV and high-voltage (HV) power systems to enhance power flow management and system stability, particularly in systems with distributed generation like wind and solar power [19]. The DAB converter consists of two H-bridges and a high-frequency transformer in between. The high-frequency transformer and switching devices reduce the size and weight of passive magnetic components [20]. One H-bridge converts the input voltage to a high-frequency AC voltage, while the other H-bridge converts back to the output DC desired voltage level.

The DAB operates in three modes: current source, voltage source, and power-sharing. It maintains a continuous flow of current to the load in the current source mode. It keeps the load terminals at a constant voltage in the voltage source mode. In the power-sharing mode, it distributes power between the two input and output sources [19]. The DAB converter offers attractive features such as a wide voltage gain range, zero voltage switching (ZVS) capability, and bidirectional power flow control [21].

### B. Benefits of Dual Active Bridge

DABs offer several advantages that make them superior to other DC-DC converters, including [22]:

1- They have improved power flow regulation in systems with high levels of distributed generation, such as wind and solar power. DABs enable better power output and consumption balancing by utilizing two parallel converters.

2- They have increased system stability by mitigating the impact of converter failures. A single converter failure in traditional systems can lead to a system-wide cascaded failure. However, with DABs, even if one converter malfunctions, the other can continue operating, ensuring system functionality and reducing the risk of outages.

Overall, DABs have significant potential benefits for medium-voltage and high-voltage power systems, including enhanced power flow management and increased system stability. They are particularly suitable for systems with high levels of distributed generation, such as wind and solar power, and can contribute to grid reliability and outage prevention.

### C. Dual Active Bridge-High Frequency Transformer

Conventional LFTs are commonly used in distribution grids to provide galvanic isolation and voltage scaling between different voltage levels. However, LFTs are bulky and lack control options due to their low operating frequency of 50Hz or 60Hz [15]. The concept of electronic transformers for grid-level applications took nearly three decades to gain acceptance, despite being patented in the 1970s. SSTs are recommended for traction systems as they offer potential size and weight reduction and improved efficiency, especially for distributed traction systems. The main idea behind SSTs is to use medium- to high-frequency isolation to transform voltage, potentially reducing volume and weight compared to traditional power transformers [23]. The current power supply design trend is miniaturization, achieved by increasing the switching frequency. However, there is a limit to the frequency increase as it leads to higher core and copper losses, reducing efficiency and causing temperature rise [24].

## III. THERMAL MANAGEMENT IN POWER SYSTEM

Thermal management and effective cooling systems are vital for industrial systems, particularly in electronic and electric applications. Maintaining optimal temperature and humidity is crucial for the equipment's safe and efficient operation. Given the significance of power electronics technologies, such as electric mobility, numerous researchers are actively studying advanced thermal management methods for various systems. For instance, Nadjahani et al. [25] explore cooling strategies for data centers, including free cooling, liquid cooling, two-phase technologies, and building envelope considerations. Qian et al. [26] review thermal management approaches for insulated gate bipolar transistors.

Efficient cooling systems are crucial for electrical transformers and high-power electronic devices to prevent thermal issues and maintain optimal performance. Heat transfer through convection, conduction, and radiation plays a key role in dissipating heat. The performance of the thermal management system is vital for the overall system design and should consider the electrical and thermal characteristics of all components [27].

McGlen et al. [28] present an overview of thermal management techniques for high-power electronic devices. They discuss the pros and cons of each method, provide insights on performance optimization, and guide the selection of the most suitable technique for specific applications.

In the following, the primary thermal management techniques used to address the thermal issues that can arise from high-power electronic devices are introduced, and some examples of the research done in this area to improve their efficiency have been presented.

*A. Forced-Air Cooling*

Using heat sinks, thermal management in embedded systems is straightforward. They are compatible globally and self-powered. Heat sinks prevent combustion by transferring heat from a high-temperature device, such as an embedded system, to a superior medium, such as air [29].

The efficacy of heat sinks with forced air-cooling fans or blowers is enhanced [30]. Heat sinks have been investigated in a variety of methods. He et al. [29] investigate miniature heat sinks for electronic device thermal control and temperature uniformity. They examine micro heat sink varieties, design, optimization, and issues. Al-damook [31] investigates how perforated, slotted, and serrated pin heatsinks reduce CPU temperature and fan power. Increasing pin perforations and implementing slotted/notched pinned heat sinks may increase CPU temperature, fan power consumption, and heat transfer rate, according to the Nusselt number.

Heat sinks with fins enhance surface area and heat transfer. Fin geometry is essential. Small electronic devices with finned heat sinks were evaluated under natural and forced convection conditions [32]. The use of straight, pin, and undulating fins was evaluated. Under natural convection, the heat sink with undulating fins had the highest heat transfer coefficient. In contrast, the heat sink with straight fins had the highest heat transfer coefficient under forced convection.

*B. Liquid Cooling*

Electronic systems manage heat well using liquid cooling. It utilizes water to chill electronic devices. Liquid cooling has superior heat dispersion and lower fan velocities than air cooling, reducing energy consumption and disturbance [26]. Researchers discovered that liquid cooling solutions could resolve thermal problems in high-energy-density electric vehicle batteries. Liquid cooling reduces battery temperatures and offers viable design options for battery cells [27].

Liquid cooling systems may benefit the power electronics in fuel-cell electric vehicles. [33] Park analyzed a liquid cooling solution for power electronics in fuel-cell electric vehicles. According to the investigation, the cooling system effectively regulates the temperature of electrical electronics. The mechanism efficiently and uniformly distributes heat

The design and operation of power electronics require thermal control. Overheating may result in component degradation and device failure [34]. Liquid cooling is a prevalent method for removing heat from power electronics. It outperforms air conditioning technologies and is indispensable for thermally demanding applications such as aviation power electronics [35]. Liquid cooling circulates water or a water-based solution over power electronics components in a closed cycle. This fluid absorbs heat from components, conveys it to a heat exchanger, and dissipates it in a refrigeration system. The chilled fluid is recirculated to maintain cooling effectiveness [34].

*C. Immersion Cooling*

Immersion cooling involves submerging components in a liquid coolant with superior thermal conductivity to air. Coolant transfers heat from components to a radiator or heat exchanger efficiently. This technology outperforms air conditioning via blowers and heat sinks. Mudawar [36] states immersion cooling enhances reliability, power density, thermal resistance, efficiency, and power consumption. According to the research, immersion cooling enhances the efficacy of high-power electrical devices. Birbarah et al. [37] compare water immersion cooling for high-power density electronics to air cooling in experiments and simulations. Immersion cooling with water enhances cooling efficiency, power density, compactness, size, weight, and reliability.

There exists both single-phase and two-phase immersion refrigeration. Using liquid coolants, single-phase immersion cooling transfers heat without phase shifts. Two-phase immersion cooling uses liquid-vapor coolants to exchange heat [38]. However, liquid refrigeration is not without its drawbacks. These include design complexity, installation and maintenance requirements, increased costs, and potential system or environmental damage caused by leakage. Barnes and Tuma [39] assessed coolant choice, cooling system design, corrosion, and contamination in liquid immersion cooling for power electronics systems.

*D. Direct Liquid Cooling*

This technique involves direct contact between the liquid coolant and the heated components, such as processors, resulting in a more effective heat transfer than air-cooling systems. It offers several advantages over conventional methods, including improved thermal stability and increased component reliability [40]. The close proximity of the liquid to the components expedites heat transfer and improves temperature regulation. Direct Liquid Cooling is employed extensively in high-performance computing systems, data centers, and power electronics applications [40, 41]. This method can be implemented in single-phase or two-phase liquid cooling systems, which will be discussed in the following.

*E. Two-Phase Liquid Cooling*

This method circulates a refrigerant liquid near its saturation point through a cold plate. As the liquid boils, it absorbs heat from the electronics and stores it as latent heat as it vaporizes. The heat is then transferred to a condenser and released into the atmosphere.

The two-phase cooling system achieves a greater heat transfer capacity by utilizing the latent heat of the coolant fluid. This makes it a superior option for cooling high-power components and reduces thermal stress risk, ultimately improving their dependability. Salamon et al. [42] conducted experiments to evaluate a two-phase liquid cooling system's thermal resistance, pressure drop, and flow rate under various operating conditions. The outcomes indicate that the system provides effective cooling with minimal thermal resistance and pressure decrease. The compact and lightweight cooling system makes it appropriate for applications with limited space, as it efficiently transmits component heat.

Two-phase liquid cooling necessitates complex equipment. Due to its design and installation complexity, it is more expensive than air conditioning systems. The closed-loop circuit's coolant fluid must be consistently cleansed and inspected. Two-phase refrigeration systems are most vulnerable to leakage caused by closed-loop pressure. Environmental hazards necessitate safety measures. [42].

### F. Single-Phase Liquid Cooling

Commonly employed to evacuate heat from power electronics components is single-phase liquid cooling. A single-phase fluid, such as water or a water-glycol mixture, is circulated through a closed-loop system. The fluid absorbs heat from the components as it passes through them and then transfers it to a heat exchanger. Then, the liquid is recirculated to the power electronics [40].

This simple technique efficiently cools power electronics, offering a high heat transfer rate, improving performance and durability. It is straightforward to implement and requires only a pump, heat exchanger, and cooling tower. Lindh et al. [41] demonstrate the viability of direct liquid cooling in low-power electrical devices by reducing component temperature and enhancing efficiency and dependability compared to air cooling. In addition to reducing the size and weight of the device, the cooling system's efficiency and compactness provide environmental benefits by reducing the need for cooling blowers and air conditioning. Khalaj and Halgamuge [43] provide an exhaustive analysis of thermal management techniques in data centers, highlighting the effectiveness of liquid cooling for high-density configurations.

However, single-phase liquid refrigeration presents particular difficulties and limitations. The coolant fluid's boiling temperature limits the cooling system's efficacy by imposing a maximum component operating temperature. In addition, these systems are susceptible to contamination and corrosion, which can reduce their effectiveness and longevity. Shia et al. [44] examined the corrosion of Copper, Aluminum, and Stainless-Steel cold plates in electronic systems when exposed to a propylene glycol/water coolant. Stainless-Steel cold plates demonstrated the highest corrosion resistance, followed by copper cold plates, while Aluminum cold plates experienced the most severe corrosion [41].

## IV. DESIGN OF THE COLD PLATE FOR THE PRIMARY AND THE SECONDARY SIDE OF THE POWER CONVERTER

### A. Over View

The cooling system for the power converter has been designed and simulated using ANSYS Fluent 2022 R1 and water-cooling technology. This study examined various system configurations, such as cold plate materials, coolant inlet velocities, and channel shapes. The effects of these three characteristics have also been investigated. All simulations have been conducted in a Steady-State system. The Primary Side contains six 10KV half-bridge SiC MOSFET modules, while Secondary Side has only two. This module is composed of numerous layers of diverse materials. The 3D model has been designed using Design Modeler software, which is a part of ANSYS Fluent. Each component contains two switches containing three chips. Thus, each SiC MOSFET module contains six dies that generate heat. As shown in Fig. 1. three devices are located on both sides of the cold plate on the Primary Side.

Fig. 2. Shows the schematic of this system, which is a $700\ kW$ system connected to a $13\ kV\ DC$ in the input and can generate $7.2\ kV\ DC$ in the output. The RMS current that flows through each switch as well as the anti-parallel diode is measured to estimate the conduction and switching losses. Table. 1. shows the loss each device generates on both the Primary and Secondary Sides [13].

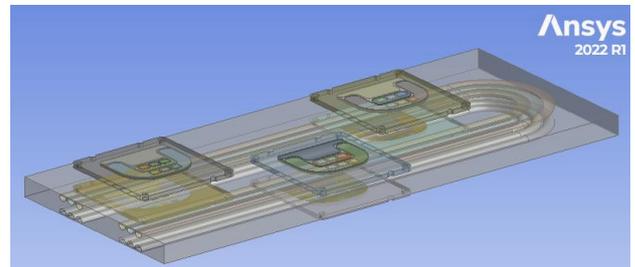

Fig. 1. 3D model of a double-sided cold plate for the Primary Side

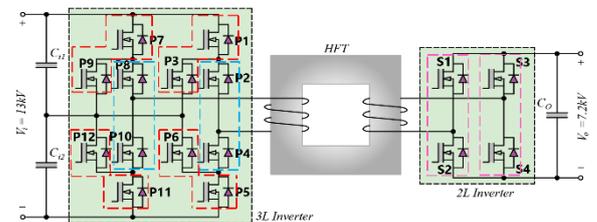

Fig. 2. SiC MOSFET half-bridges configuration to create an ANPC-2L converter

Table. 1. Calculated switching losses for the Primary and the Secondary side

| Switch Pair | P1, P7 | P2, P8 | P3, P9 | P4, P10 | P5, P11 | P6, P12 | S1, S3 | S2, S4 |
|---|---|---|---|---|---|---|---|---|
| **Loss** | 102 W | 107.16 W | 84.88 W | 91.1 W | 89.6 W | 85.16 W | 211.4 W | 127.18 W |

## B. Cooling System design for the Primary Side of the Power Converter

The "body transformation" feature of Design Modeler duplicated one device into six (three on each side). This device was created using precise layer measurements. The cold plate and its channels were then engraved. The initial dimensions of the cold plate were $480 * 190 * 18\ mm$. Before performing any other simulations, the mesh independence simulations must be performed first. By conducting these simulations, we can ensure that increasing the number of elements and decreasing the mesh size in the future will not significantly impact the ultimate outcomes. This investigation can save time by reducing the duration of simulations and improving the results' accuracy.

Fig. 3. demonstrates that as the number of elements increases, the maximum temperature decreases, but the change is negligible after approximately 7 million elements. Therefore, one of the MESHs with nearly 7 million elements was selected, and simulations for each model were conducted using this MESH.

The next step was to determine which channel type provides superior performance and system compatibility. Thus, rectangular and semicircular channels have been selected. The cold plate is equipped with two layers of channels, one for the top and one for the bottom. The initial dimensions of the rectangular channels were set to $2\ mm$ by $10\ mm$, and it was presumed that their total surface area was identical. Consequently, based on the total surface areas of the rectangular channels calculated with Eq. 1, the radius of each semicircular channel has been calculated using Eq. 2, assuming there are two rows of channels, each with three channels. This calculation yielded a radius of 4.6mm for every semicircular channel.

$$S = PLN \qquad (1)$$

$$S = (\pi r + 2r)LN \qquad (2)$$

where $S$ is the total surface area, $L$ is the average length of channels, $P$ is the perimeter of the inlet area, $N$ is the number of channels, and $r$ is the radius of one channel.

After modifying and finalizing the model and applying the MESH and all boundary conditions for the simulation, the type of simulation method, based on the flow type, needed to be chosen. Based on the Reynolds number at 2500, which is the transmission point from the Laminar flow to the Turbulent flow, the velocity of the fluid has been calculated using Eq. 3. These calculations gave us the velocity of $0.44\ m/s$ for the coolant inside of the semicircular channels and $0.75\ m/s$ for the coolant inside of the rectangular channels at the transmission point. These values of velocities will show that for all values below these numbers, the fluid movement will be Laminar, and for all values above these numbers, the fluid movement will be in Turbulent flow.

$$Re = \frac{\rho v D_h}{\mu}, \qquad (3)$$
$$D_h = \frac{4A}{P},$$

where $\rho$ is the viscosity, v is the velocity, and µ is the dynamic viscosity of the coolant, which is Water, and $D_h$ is the hydraulic diameter of the channels.

The software can handle both the Laminar and the Turbulent flow by choosing K Epsilon as a simulation method in ANSYS Fluent. Hence, there is no need to choose the Laminar method for the simulation when the fluid is in Laminar flow.

The findings of Fig. 4 show that the cold plate with semicircular channels operates better and performs much better than the cold plate with rectangular channels. The maximum temperature of the module using the semicircular channels for the cold plate was $82.4\ °C$, while it was $101.3\ °C$ for the module on the cold plate with rectangular channels. So, the design of the cooling system proceeded using semicircular channels with a radius of 4.6mm for each side of the cold plate.

The following step was to determine which material would perform best as the cold plate. Consequently, three materials have been selected: Copper, Aluminum, and Stainless-Steel. Copper has superior thermal conductivity compared to the other two, which led to selection of these three materials. Stainless steel, on the other hand, has superior corrosion resistance compared to the others. Therefore, after completing all simulations, the advantages and cons of each material have been evaluated, and the optimal solution has been selected. As expected, Fig. 5 demonstrates that the cold plate made of Copper has superior efficacy to the other materials due to its increased thermal conductivity. Although Stainless Steel has a much higher corrosion resistance than Copper, the utmost temperature it can provide for the module is high enough to disqualify it as a cold plate material. Copper was chosen as the cold plate because the maximum temperatures for the device that these cooling systems have provided for the modules at a velocity of 1.1 m/s are $122.33\ °C$ with Stainless-Steel as the cold plate, $86.18\ °C$ with Aluminum as the cold plate and $82.4\ °C$ with Copper as the cold plate.

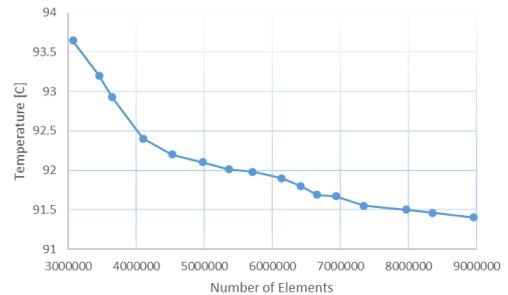

Fig. 3. Maximum Temperature of the module using a different number of elements for the MESH.

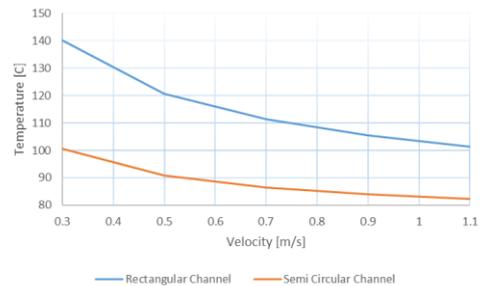

Fig. 4. Maximum temperature of the module at different velocities for two cold plates using two different channel shapes

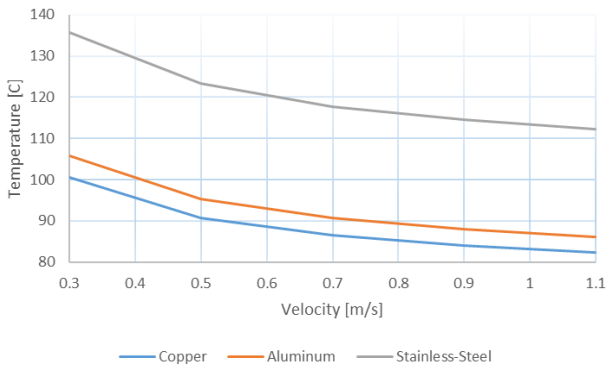

Fig. 5. Maximum temperature of the module at different velocities for three cooling systems using three different materials as the cold plate

The chosen material and channel configurations at a particular velocity worked well for the cooling system, but the total weight of the cold plate was a significant issue. It weighed close to 13 $kg$, excluding the weight of the coolant contained within the channels. It was possible to reduce the weight of the cold plate by reducing the size of the channels while maintaining the same system performance. Assuming that the channels' total surface area remains constant, their radius decreases as the number of channels increases. By increasing the number of channels on each side of the cold plate to 4, 5, and 6 using Equations 1 and 2, the channels' radius decreased from 4.6 $mm$ to 2.3 $mm$. The 3D model was modified and altered to proceed with the simulation based on the new number of channels and their radius.

By increasing the number of channels and reducing their dimensions, as depicted in Fig. 6. the maximal temperature of the modules has increased. However, at a velocity of 1.1 $m/s$, the cold plate with six channels on each side can provide the module with a maximal temperature 1.89 °C higher than the cold plate with three channels on each side which is a negligible difference.

By employing six semicircular channels with a radius of 2.3 $mm$ on each side of the cold plate, the plate's thickness has been reduced from 18 $mm$ to 7.6 $mm$, and its vacant spaces have been eliminated. Without the coolant in the channels, the weight of the cold plate decreased from 13 $kg$ to 3.76 $kg$. Fig. 7. depicts the finalized model for the cold plate of the Primary Side of the power converter.

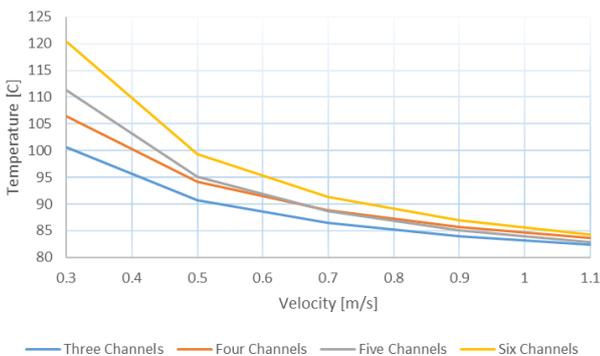

Fig. 6. Maximum temperature of the module at different velocities for cold plates with different numbers of channels

The simulation result for the finalized cold plate, depicted in Fig. 8. indicates that the system's maximum temperature increased by 1.38 °C. Although the 10 $kV$ module can operate at temperatures as high as 170 °C, it was anticipated that a cooling system would be developed to maintain the maximum temperature of these modules below 135 °C. The final design of the cold plate for the Primary Side of the power converter can be derived from this model. Fig. 9. and Fig. 10. display the pressure drop of the coolant within the channels, which is 16791.3 $Pa$ (167.91 $mBar$), and the temperature of the coolant within the channels, which is between 49 °C and 56 °C.

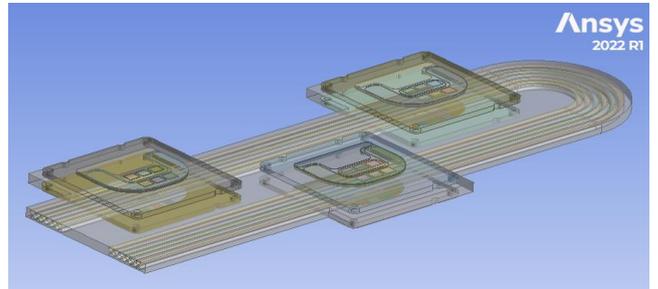

Fig. 7. 3D model of the finalized cold plate for the Primary Side of the power converter

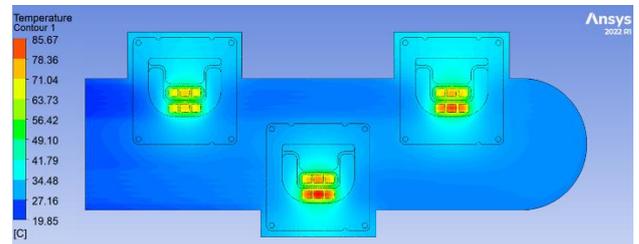

Fig. 8. Temperature distribution of the finalized cold plate for the Primary Side of the power converter

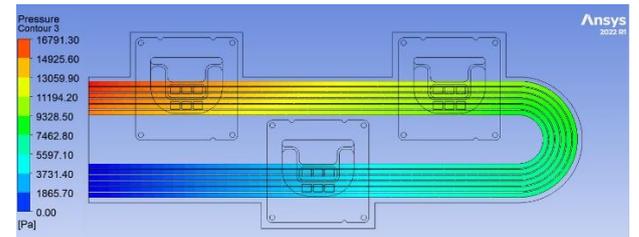

Fig. 9. Pressure drops of the coolant inside of the channels of the finalized cold plate for the Primary Side of the power converter

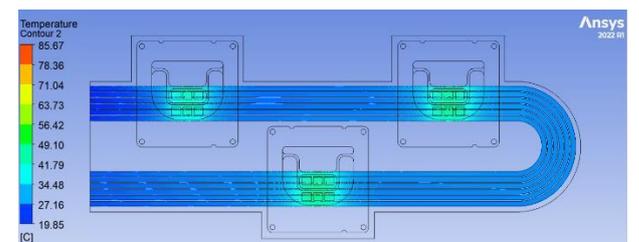

Fig. 10. Temperature distribution of the coolant inside of the channels of the finalized cold plate for the Primary Side of the power converter

## C. COLD PLATE FOR SECONDARY

Since the power loss of the modules on the Secondary Side was greater than on the Primary Side, it was determined that a single-sided cold plate would improve the cooling system's performance. Consequently, the model was modified to include a cold plate with the dimensions $330 * 205 * 7.6\ mm$, two rows of channels with six semicircular channels having a radius of $2.3\ mm$, and two $10\ kV$ Modules on top of the cold plate. Due to their higher loss, the simulation began with an initial inlet velocity of $1.1\ m/s$.

The simulation results showed that the initial maximum temperature that this cooling system provided for the module was $144.93\ °C$ which is a much higher temperature compared to the Primary Side. The inlet velocity of the coolant increased to $1.4\ m/s$ for the second attempt, which caused the maximum temperature to decrease to $142.04\ °C$. In all models, the distance which has been set between the channels and the surface of the cold plate was $1\ mm$. By reducing this distance from $1\ mm$ to $0.5\ mm$, an impressive drop in the maximum temperature of the device has been accrued, and it became $136.86\ °C$. Fig. 11. shows that increasing the velocity from $1.4\ m/s$ to $2.9\ m/s$ reduced the maximum temperature of the device from $136.86\ °C$ to $131.58\ °C$ which was the maximum temperature of the module that this cold plate provided for the Secondary Side of the power converter.

By using this cooling system, the module's highest temperature remains acceptable. Consequently, this model can serve as the final cold plate design for the Secondary Side of the power converter. Fig. 12. to Fig. 14. Display the temperature distribution of the cold plate, the pressure decrease of the coolant within the channels, which is $16791.3\ Pa$ ($167.91\ mBar$), and the temperature of the coolant within the channels, which is between $49\ °C$ and $56\ °C$. At a velocity of $2.9\ m/s$, a copper cold plate with two rows of channels, each with six semicircular channels with a radius of $2.3\ mm$, performed optimally.

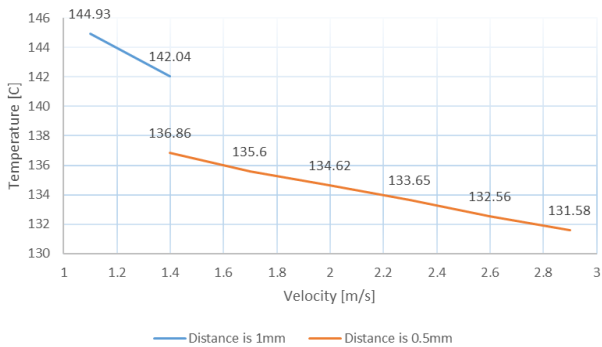

Fig. 11. Maximum temperature of the module at different velocities for cold plates with different numbers of channels

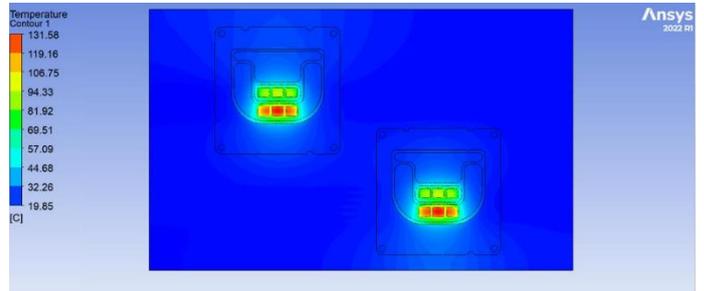

Fig. 12. Temperature distribution of the finalized cold plate for the Primary Side of the power converter

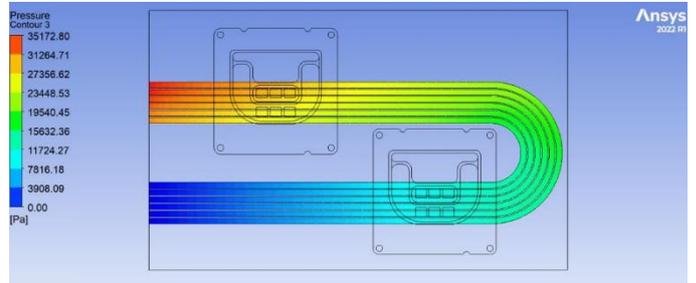

Fig. 13. Pressure drops of the coolant inside of the channels of the finalized cold plate for the Primary Side of the power converter

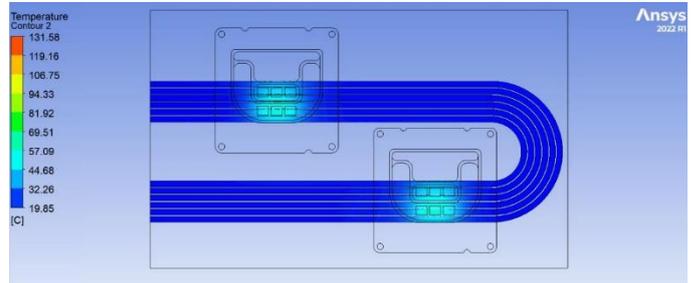

Fig. 14. Temperature distribution of the coolant inside of the channels of the finalized cold plate for the Primary Side of the power converter

## V. CONCLUSION

ANSYS Fluent was utilized to simulate a cooling system for the Primary and Secondary Sides of the power converter. The simulation investigated the effects of channel size, geometry, cold plate material, and coolant inlet velocity on system efficiency. Water was used as the coolant in the simulation, along with a pressure-based solver and a k-epsilon realizable model. By utilizing these two cold plates for the Primary and Secondary Sides of the power converter, the maximum temperature of the modules was reduced from $140.2\ °C$ to $84.29\ °C$ on the Primary Side and from $144.93\ °C$ to $131.58\ °C$ on the Secondary Side.

This research shows the benefits of utilizing simulation tools like ANSYS Fluent to enhance electrical device cooling system designs. Simulation findings improve performance, reliability, and efficiency and save time and money compared to physical prototypes. Our team is building and testing cold plates to improve transformer water-cooling systems. This repeated process validates and solidifies our simulation findings and results, confirming the system's viability and dependability for varied transformer applications.